\documentstyle[aps,prb,twocolumn]{revtex}

\begin{document}
\title{Self-organization of vortices in type-II superconductors during magnetic
relaxation}
\author{R. Prozorov}
\address{Loomis Laboratory of Physics, University of Illinois at Urbana-Champaign,\\
1110 W. Green St., Urbana, IL 61801, U. S. A.}
\author{D. Giller}
\address{Institute of Superconductivity, Department of Physics, Bar-Ilan University,\\
52900 Ramat-Gan, Israel.}
\date{submitted to Phys. Rev. B - 8 November 1998; accepted - 5 March 1999}
\maketitle

\begin{abstract}
We revise the applicability of the theory of self-organized criticality
(SOC) to the process of magnetic relaxation in type-II superconductors. The
driving parameter of self-organization of vortices is the energy barrier for
flux creep and not the current density. The power spectrum of the magnetic
noise due to vortex avalanches is calculated and is predicted to vary with
time during relaxation.
\end{abstract}


\section{Introduction}

The magnetic response of hard type-II superconductors, in particular
magnetic flux creep, is a timely issue in contemporary research (see for
review \cite{blatter,brandt,yeshurun}). In early 60s a very useful model of
the critical state was developed to describe magnetic behavior of type-II
superconductors \cite{bean,anderson,campbell,gennes}. One of the
distinguishing features of this behavior, observed experimentally, is that
the density of flux lines varies across the whole sample. This model of the
critical state remains in use, even though a significant progress has been
made in understanding the particular mechanisms of a magnetization and creep
in type-II superconductors \cite{blatter,brandt,yeshurun,campbell}. It has
also been noted that the magnetic flux distribution in type-II
superconductors is, in many aspects, similar to a sandpile formed when, for
example, sand is poured onto a stage \cite{campbell,gennes,tang}. When a
steady state is reached the slope of such a pile is analogous to the
critical current density $j_{c}$ of a superconductor. Study of the dynamics
(i. e. sand avalanches) of such strongly-correlated many-particle systems
has led to a development of a new concept, called self-organized criticality
(SOC), proposed originally by Bak and co-workers \cite{bak}. Tang first
analyzed direct application of SOC to type-II superconductors \cite{tang}.
Later numerous studies significantly elaborated on this topic \cite
{vinokur,soc,field,pan,bonabeau,olson}.

In practice, especially in high-T$_{c}$ superconductors, persistent current
density $j$ in the experiment is much lower than the critical current
density $j_{c}$ due to ''giant'' flux creep \cite{yeshurun}. The concept of
SOC is strictly applied only to the critical state $j=j_{c}$ \cite
{field,pan,bonabeau} and it describes the system dynamics {\em towards} the
critical state. Nevertheless, it is tempting to analyze magnetic flux creep
in type-II superconductors during which the system moves {\em out} of the
critical state, in a SOC context, because thermal activation can trigger
vortex avalanches \cite{field,pan,bonabeau}. However, it was found that
modifications of the relaxation law due to vortex avalanches are minor and
can hardly be reliably distinguished in the analysis of experimental data.
Furthermore, flux creep {\em universality} has been analytically
demonstrated in the elegant paper by Vinokur {\it et al.} \cite{vinokur}.
Universality of the spatial distribution of the electric field during flux
creep has also been found by Gurevich and Brandt \cite{gurevich}. The direct
application of SOC to the problem of magnetic flux creep thus meets a number
of serious general difficulties. It is clear that critical scaling (power
laws for vortex-avalanche lifetimes and size distributions) observed in the
vicinity of the critical state \cite{bak,tang} must change during later
stages of relaxation due to a time-dependent (or current-dependent) balance
of the Lorentz and pinning forces.

In this paper we propose a new physical picture of self-organization in a
vortex matter during magnetic flux creep in type-II superconductors. In this
approach the driving parameter is the energy barrier for magnetic flux creep
rather than the current density. We show that notwithstanding its minor
influence on the relaxation rate, self-organized behavior may be observed by
measuring magnetic noise during flux creep.

\section{Barrier for magnetic flux creep as the driving parameter of
self-organization}

We consider a long superconducting slab infinite in the $y$ and $z$
directions and having width $2w$ in the $x$ direction. The magnetic field is
directed along the $z$ axis. In this geometry, the flux distribution is
one-dimensional, i.e., ${\bf B}\left( {\bf r},t\right) =\left( 0,0,B\left(
x,t\right) \right) $. As a mathematical tool for our analysis we use a well
known differential equation for flux creep \cite{blatter,brandt,anderson}: 
\begin{equation}
\frac{\partial B}{\partial t}=-\frac{\partial }{\partial x}\left( Bv_{0}\exp
\left( -U\left( B,T,j\right) /T\right) \right)  \label{dbdt}
\end{equation}

\noindent Here $B$ is the magnetic induction, $v=v_{0}\exp \left( -U\left(
B,T,j\right) /T\right) $ is the mean velocity of vortices in the $x$
direction and $U\left( B,T,j\right) $ is the effective barrier for flux
creep. Note that we adopt units with $k_{B}=1$, thus energy is measured in
kelvin. Since in our geometry $4\pi M=%
%TCIMACRO{\tint}%
%BeginExpansion
\textstyle\int%
%EndExpansion
\limits_{V}\left( B-H\right) dV$ we get for the mean volume magnetization $%
m=M/V$ from Eq. \ref{dbdt}: 
\begin{equation}
\frac{\partial m}{\partial t}=-A\exp \left( -U\left( H,T,j\right) /T\right)
\label{dmdt}
\end{equation}
where $A\equiv Hv_{0}/4\pi w$.

It is important to emphasize that we do not modify the pre-exponent factor $%
Bv_{0}$ of Eq. \ref{dbdt} or $A$ of Eq. \ref{dmdt}, as suggested by previous
works on SOC (see e.g. \cite{tang}). Such modifications result only in
logarithmic corrections to the effective activation energy, and they may be
omitted in a flux creep regime \cite{field,pan,bonabeau}. Instead, we
concentrate on the details of the spatial behavior of flux creep barrier $%
U\left( x\right) $, as analyzed in detail in our previous work \cite
{burlachkov}. In that work Eq. \ref{dbdt} was solved numerically and
semi-analytically for different situations. We emphasize that, in general,
the barrier for flux creep depends on magnetic field $B$, and persistent
current density $j\left( x\right) $ is not uniform across the sample (see
Fig. 1). Thus, $j$ cannot be used as a driving parameter for a SOC model.
Instead the relevant parameter is $U$, which stays constant across the
sample. Also, since experiments on magnetic relaxation are usually carried
out at constant temperature and at high magnetic field, we can assume $%
U\left( B,T,j\right) =U\left( j\right) $. Central results of Ref. \cite
{burlachkov} are shown in Fig. 1 using a ''collective creep'' - type
dependence $U\left( j\right) =U_{0}\left( B/B_{0}\right) ^{n}\left( \left(
j_{c}/j\right) ^{\mu }-1\right) $, with $n=5$ and $\mu =1$as an example, see
also Eq. \ref{griessen} below (other models are analyzed in Ref. \cite
{burlachkov} as well and produce essentially similar results). Filled
squares in Fig. 1 represent the distribution of the magnetic induction $%
B\left( x\right) /H$ at some late stage of relaxation (so that $j<j_{c}$),
the solid line represents the normalized current density profile (note that $%
j_{c}$ is constant across the sample), and open circles show the profile of
the effective barrier for flux creep $U\left( x\right) /T$. All quantities
are calculated numerically from Eq. \ref{dbdt}. The important thing to note
is that the energy barrier $U\left( x\right) $ is nearly independent of $x$,
so that its maximum variation $\delta U$ is of order of $T$. As also shown
from general arguments \cite{burlachkov}, such behavior means that the
fluxon system organizes itself to maintain a uniform distribution of the
barrier $U$ across the sample.

The vortex avalanches are introduced in an integral way. An avalanche of
size $s$ causes a change in the total magnetic moment $\delta M\equiv s$.
This change is equivalent to a change of the average current density $\delta
j=\delta M\gamma =\gamma s$, where $\gamma =2c/wV$. If the barrier for flux
creep is $U\left( j\right) $, then the variation of current $\delta j$ leads
to a variation of the energy barrier

\begin{equation}
\delta U=\left| \frac{\partial U}{\partial j}\right| \delta j=\gamma \left| 
\frac{\partial U}{\partial j}\right| s  \label{dU}
\end{equation}

As mentioned above, maximum fluctuation in the energy barrier $\left| \delta
U\right| _{\max }$ is of order of $T$ in the creep regime ($\delta U<<U$).
Any fluctuation $\delta U$ larger than $T$ is suppressed before it arrives
to the sample edge due to exponential feedback of the local relaxation rate,
which is proportional to $\exp \left( -U/T\right) $, (Eq. \ref{dbdt}). This
means that only fluctuations $\delta U\leq T$ can be observed in global
measurements of the sample magnetic moment. Thus,

\begin{equation}
s_{m}=\frac{T}{\gamma \left| \frac{\partial U}{\partial j}\right| }\propto VT
\label{sm}
\end{equation}

\noindent where we denote as $s_{m}$ the maximum possible avalanche, which
depends on time via $\partial U/\partial j$. It is worth to note that Eq. 
\ref{sm} gives the correct dependence of $s_{m}$ on the system size and on
temperature. It is clear that in a finite system the largest possible
avalanche must be proportional to the system volume. Since it is thermally
activated, it is proportional to temperature $T$, consistent with our
derivation. The characteristic time-dependent upper cut-off of the avalanche
size was experimentally observed by Field {\it et al.} \cite{field} who
studied magnetic noise spectra at different magnetic field sweep rates, i.e.
at different time windows of the experiment.

Our central idea is that in the vicinity of $j_{c}$ the system of fluxons,
indeed, exhibits self-organized {\em critical} behavior, as initially
proposed by Tang \cite{tang}. During flux creep, it maintains itself in a
self-organized, however {\em not critical} state in the sense that it cannot
be described by the critical scaling. The self-organization manifests itself
by the appearance of almost constant across the sample $U$. Avalanches do
not vanish, but there is a constrain on the largest possible avalanche, see
Eq. \ref{sm}. Importantly, $s_{m}$ depends upon current density and, as we
show below, decreases with decrease of current (or with increase of time),
so their relative importance vanishes.

In order to calculate physically measured quantities let us derive the time
dependence of $s_{m}$ assuming a very useful generic form of the barrier for
flux creep, introduced by Griessen \cite{griess}. 
\begin{equation}
U\left( j\right) =\frac{U_{0}}{\alpha }\left[ \left( \frac{j_{c}}{j}\right)
^{\alpha }-1\right]  \label{griessen}
\end{equation}
This formula describes all widely-known functional forms of $U\left(
j\right) $ if the exponent $\alpha $ attains both negative and positive
values. For $\alpha =-1$ Eq. \ref{griessen} describes the Anderson-Kim
barrier \cite{anderson}; for $\alpha =-1/2$ the barrier for plastic creep 
\cite{abulafia} is obtained. Positive $\alpha $ describes collective creep
barriers \cite{blatter}. In the limit $\alpha \rightarrow 0$ this formula
reproduces exactly logarithmic barrier \cite{zeldov}. An activation energy
written in the form of Eq. \ref{griessen} results in an ''interpolation
formula'' for flux creep \cite{blatter} if the logarithmic solution of the
creep equation $U\left( j\right) =T\ln (t/t_{0})$ is applied \cite
{geshkenbein} (for $\alpha \neq 0$): 
\begin{equation}
j\left( t\right) =j_{c}\left( 1+\frac{\alpha T}{U_{0}}\ln \left( \frac{t}{%
t_{0}}\right) \right) ^{-\frac{1}{\alpha }}  \label{jinterp}
\end{equation}
For $\alpha =0$, a power-law decay is obtained: $j\left( t\right)
=j_{c}\left( t_{0}/t\right) ^{n}$, where $n=T/U_{0}$.

Using this general form of the current dependence of the activation energy
barrier, we obtain from Eq. \ref{sm} 
\begin{equation}
s_{m}\left( j\right) =\frac{Tj}{\gamma U_{0}}\left( \frac{j}{j_{c}}\right)
^{\alpha }  \label{smjgries}
\end{equation}
and 
\begin{equation}
s_{m}\left( t\right) =\frac{Tj_{c}}{\gamma U_{0}}\left( 1+\frac{\alpha T}{%
U_{0}}\ln \left( \frac{t}{t_{0}}\right) \right) ^{-\left( 1+\frac{1}{\alpha }%
\right) }.  \label{smtgries}
\end{equation}
As we see, the upper limit for the avalanche size decreases with the
decrease of current density or with the increase of time for all $\alpha >-1$%
. For $\alpha <-1$ the curvature 
\begin{equation}
\frac{\partial ^{2}U}{\partial j^{2}}=\frac{\left( \alpha +1\right) }{j^{2}}%
U_{0}\left( \frac{j_{c}}{j}\right) ^{\alpha }  \label{curvature}
\end{equation}
is negative and largest avalanche does not change with current, but is
limited by its value at criticality). In this case, self-organized {\em %
criticality} describes the system dynamics down to very low currents. On the
other hand the Kim-Anderson barrier must be always relevant when $%
j\rightarrow j_{c}$ \cite{blatter}, thus our model produces a correct
transition to a self-organized critical state at $j=j_{c}$. In practice,
most of the observed cases obey $\alpha \geq -1$ and $s_{m}$ decreases with
decrease of current density (due to flux creep).

\section{Avalanche distributions and the power spectrum}

Before starting with calculation of the power spectrum of the magnetic flux
noise due to flux avalanches, let us stress that the time dependence of $%
s_{m}$ is very weak (logarithmic, see Eq. \ref{smtgries}). This allows us to
treat the process of the flux creep as quasi-stationary, which means that
during a short time, as required for the sampling of the power spectrum,
current density is assumed to be constant. In more sophisticated experiments 
\cite{field} the external field can be swept with the constant rate, which
insures that the current density does not change, although $j<j_{c}$.
Actually, constant sweep rate fixes a certain time window of the experiment $%
t/t_{0}\propto 1/\left( \partial H/\partial t\right) $. Thus, decreasing the
sweep rate allows the noise spectra to be studied at effectively later
stages of the relaxation. {\em \ }

Once an avalanche is triggered by a thermal fluctuation, its subsequent
dynamics is governed only by interactions between vortices for which motion
is {\it not} due to thermal fluctuations. Thus, we expect same relationship
between the avalanche lifetime $\tau $ and its size $s$ as in the case of a
sandpile: $\tau \left( t\right) \propto s^{\sigma }\left( t\right) $ and $%
\tau _{m}\left( t\right) \propto s_{m}^{\sigma }\left( t\right) $,
respectively. Using the simplified version of the distribution of lifetimes
estimated for a superconductor in a creep regime from computer simulations
by Pan and Doniach \cite{pan},

\begin{equation}
\rho \left( \tau \right) \propto \exp \left( -\tau /\tau _{m}\right) ,
\label{rho(tau)}
\end{equation}
and assuming that avalanches of size $s$ and lifetime $\tau $ contribute the
Lorentzian spectrum, 
\begin{equation}
L\left( \omega ,\tau \right) \propto \frac{\tau }{1+\left( \omega \tau
\right) ^{2}}  \label{Lom}
\end{equation}
the total power spectrum of magnetic noise during flux creep is 
\begin{equation}
S\left( \omega \right) =\int\limits_{0}^{\infty }\rho \left( \tau \right)
L\left( \omega ,\tau \right) d\tau .  \label{som}
\end{equation}
Using Eq. \ref{rho(tau)} we find:

\begin{equation}
S\left( p\right) \propto \frac{1}{2p^{2}}\left[ \cos \left( \frac{1}{p}%
\right) 
%TCIMACRO{\func{Re}}%
%BeginExpansion
\mathop{\rm Re}%
%EndExpansion
\left( 
%TCIMACRO{\func{Ei}}%
%BeginExpansion
\mathop{\rm Ei}%
%EndExpansion
\left( \frac{i}{p}\right) \right) -\sin \left( \frac{1}{p}\right) 
%TCIMACRO{\func{Im}}%
%BeginExpansion
\mathop{\rm Im}%
%EndExpansion
\left( 
%TCIMACRO{\func{Ei}}%
%BeginExpansion
\mathop{\rm Ei}%
%EndExpansion
\left( \frac{i}{p}\right) \right) \right] .  \label{spectrum}
\end{equation}
Here $p\equiv \omega \tau _{m}\left( t\right) $ and $%
%TCIMACRO{\func{Ei}}%
%BeginExpansion
\mathop{\rm Ei}%
%EndExpansion
\left( x\right) =\int\limits_{0}^{\infty }e^{-x\eta }/\eta d\eta $ is the
exponential integral. The power spectrum $S\left( \omega ,t\right) $
described by Eq. \ref{spectrum} is plotted in Fig. 2 using a solid line.
Since there an upper cutoff for the avalanche lifetime at $\tau _{m}$, the
lowest frequency which makes sense is $2\pi /\tau _{m}$. Thus, only
frequency domain $2\pi /\tau _{m}<\omega $ ($p>1$) is important. In the
limit of large $p$, the spectral density of Eq. \ref{spectrum} has a simple
asymptote: 
\begin{equation}
S\left( \omega \right) \propto \frac{\ln \left( p\right) -\gamma _{e}}{p^{2}}
\label{spectrum1}
\end{equation}
where $\gamma _{e}\approx 0.577...$ is Euler's constant. This simplified
power spectrum is shown in Fig. 2 by a dashed line. For $p>10$ this
approximation is quite reasonable. The usual way to analyze the power
spectrum is to present it in a form $S\left( \omega \right) \propto 1/\omega
^{\nu }$ and extract the exponent $\nu $ simply as $\nu =-\partial \ln
\left( S\right) /\partial \ln \left( \omega \right) $. In our case the
parameter $p=\omega \tau _{m}$ is a reduced frequency, so the exponent $\nu $
can be estimated as 
\begin{equation}
\nu =-\frac{\partial \ln \left( S\right) }{\partial \ln \left( p\right) }=2-%
\frac{1}{\ln \left( p\right) -\gamma _{e}}  \label{nu}
\end{equation}
This result is very important, since it fits quite well the experimentally
observed values of $\nu $ which were found to vary between $1$ and $2$ \cite
{field,ferrari}. As seen from Fig. 2, it is impossible to distinguish
between real $1/\omega ^{\nu }$ dependence and that predicted by Eq. \ref
{spectrum} at large enough frequencies. Remarkably, in many experiments the
power spectrum was found to deviate significantly from the $1/\omega ^{\nu }$
behavior at lower frequencies, which fits, however, Eq. \ref{spectrum}.

Using Eq. \ref{spectrum} or Eq. \ref{spectrum1} one can find the
temperature, magnetic field and time dependence of the power spectrum
substituting $p=\omega \tau _{m}=\omega s_{m}^{\sigma }$ and using values of 
$s_{m}\left( H,T,t\right) $ derived in the previous section. Specifically,
from Eq. \ref{smtgries} we obtain that any given frequency amplitude of a
power spectrum increases with time in the collective creep regime, but
saturates in the case of the logarithmic barrier and remains constant in the
case of the Kim-Anderson barrier.

In general, we emphasize that the power spectrum of the magnetic noise
during flux creep depends on time. Since parameter $p$ decreases with the
increase of time, the exponent $\nu $ becomes closer to $1$ during flux
creep. At these later stages of relaxation the effect of the avalanches is
negligible and magnetic noise is mostly determined by thermally activated
jumps of vortices with the usual (non-correlated) $1/\omega $ power
spectrum. Thus, the manifestation of the avalanche-driven dynamics during
flux creep is noise spectra with $1/\omega ^{\nu }$ and decreasing $\nu
\left( t\right) $ when sampled at different times during relaxation. This
explains the experimental results obtained by Field {\it et al.} \cite{field}%
, who measured directly vortex avalanches at different sweep rates. Those
found that the exponent $\nu $ decreased from a relatively large value of $2$
at a large sweep rate of $20\ G/\sec $ to a smaller value of $1.5$ for a
sweep rate of $1\ G/\sec $. This is in a good agreement with our model.

\section{Conclusions}

In conclusion, self-organization of vortices in hard type-II superconductors
during magnetic flux creep was analyzed. Using results of a numerical
solution \cite{burlachkov} of the differential equation for flux creep, it
was argued that the self-organized {\em criticality} describes the system
dynamics at $j=j_{c}$. During flux creep, the vortex system remains{\em \
self-organized}, but there is {\em no criticality} in the sense that there
are no simple power laws for distributions of the avalanche size, lifetime,
and for the power spectrum. The driving parameter of the self-organized
dynamics is the energy barrier $U\left( B,j\right) $ and not the current
density $j$, as proposed by previous work. Using a simple model the power
spectrum $S\left( \omega \right) $ of the magnetic noise is predicted to
depend on time. Namely, fitting $S\left( \omega \right) $ to a $1/\omega
^{\nu }$ behavior will result in a time-dependent exponent $\nu \left(
t\right) $ decreasing in the interval between $2$ and $1$.

{\sl Acknowledgments: }We acknowledge fruitful discussions with L.
Burlachkov and B. Shapiro. We thank F. Nori for critical remarks. D.G.
acknowledges support from the Clore Foundations. This work was partially
supported by the National Science Foundation (DMR 91-20000) through the
Science and Technology Center for Superconductivity, and by DOE grant
DEFG02-91-ER45439.

\newpage{}

{\Large Figure captions}

Fig.1 Results of numerical solution of Eq. \ref{dbdt} for $U\left( j\right)
=U_{0}\left( B/B_{0}\right) ^{5}\left( j_{c}/j-1\right) $ at $j<j_{c}$.
Spatial distribution of magnetic induction $B\left( x\right) /H$ (filled
squires); corresponding profile of the normalized current density (solid
line) and the corresponding profile of the effective barrier for flux creep $%
U\left( x\right) /T$ (open circles).

Fig. 2 The power spectrum $S\left( \omega ,t\right) $ described by Eq. \ref
{spectrum} (solid line) and approximate asymptotic solution (dashed line).


\begin{references}
\bibitem{blatter}  G. Blatter, M. V. Feigelman,V. B. Geshkenbein, A. I.
Larkin, and V. M. Vinokur, Rev. Mod. Phys. {\bf 66}, 1125 (1994).

\bibitem{brandt}  E. H. Brandt, Rep. Prog. Phys. {\bf 58}, 1465 (1995).

\bibitem{yeshurun}  Y. Yeshurun, A. P. Malozemoff, and A. Shaulov, Rev. Mod.
Phys. {\bf 68}, 911 (1996).

\bibitem{bean}  C. P. Bean, Phys. Rev. Lett. {\bf 8}, 250 (1962). Y. B. Kim,
C. F. Hempstead, and A. R. Strand, Physical Review 129, 528 (1963).

\bibitem{anderson}  P. W. Anderson, Phys. Rev. Lett. {\bf 9}, 309 (1962); Y.
B. Kim, C. F. Hempstead, and A. R. Strand, Phys. Rev. Lett. {\bf 9}, 306
(1962); Y. B. Kim, C. F. Hempstead and A. R. Strand, Phys. Rev. {\bf 129},
528 (1963); P. W. Anderson and Y. B. Kim, Rev. Mod. Phys. {\bf 36}, 39
(1964); M. R. Beasley, R. Labush, and W. W. Webb, Phys. Rev. {\bf 181}, 682
(1969).

\bibitem{campbell}  A. M. Campbell and J. E. Evetts, {\it ''Critical
currents in superconductors''} (Taylor Francis Ltd., London, 1972); A. I.
Larkin and Y. N. Ovchinnikov, J. Low Temp. Phys. {\bf 73}, 109 (1979); H.
Ullmaier, {\it ''Irreversible properties of type-II superconductors''}
(Springer-Verlag, Berlin, Heidelberg, New York, 1975); E. H. Brandt and M.
V. Indenbom, Phys. Rev. B {\bf 48}, 12893 (1993); E. Zeldov, J. R. Clem, M.
McElfresh, and M. Darwin, Phys. Rev. B {\bf 49}, 9802 (1994); E. H. Brandt,
Phys. Rev. Lett. {\bf 74}, 3025 (1995);

\bibitem{tang}  C. Tang, Physica A {\bf 194}, 315 (1993).

\bibitem{gennes}  P. G. de Gennes, {\it ''Superconductivity of metals and
alloys''} (W. A. Benjamin, New-York, 1966).

\bibitem{bak}  P. Bak, C. Tang, and K. Wiesenfeld, Phys. Rev. Lett. {\bf 59}%
, 381 (1987); P. Bak, C. Tang, and K. Wiesenfeld, Phys. Rev. A {\bf 38}, 364
(1988); C. Tang and P. Bak, Phys. Rev. Lett. {\bf 60}, 2347 (1988).

\bibitem{vinokur}  V. M. Vinokur, M. V. Feigelman, and V. B. Geshkenbein,
Phys. Rev. Lett. {\bf 67}, 915 (1991).

\bibitem{soc}  O. Pla and F. Nori, Phys. Rev. Lett. {\bf 67}, 919 (1991); Z.
Wang and D. Shi, Phys. Rev. B {\bf 48}, 4208 (1993); Z. Wang and D. Shi,
Phys. Rev. B {\bf 48}, 9782 (1993);

\bibitem{field}  S. Field, J. Witt, F. Nori, and X. Ling, Phys. Rev. Lett. 
{\bf 74}, 1206 (1995);

\bibitem{pan}  W. Pan and S. Doniach, Phys. Rev. B {\bf 49}, 1192 (1994).

\bibitem{bonabeau}  E. Bonabeau and P. Lederer, Physica C {\bf 256}, 365
(1996).

\bibitem{olson}  C. J. Olson, C. Reichhardt, and F. Nori, Phys. Rev. B {\bf %
56}, 6175 (1997).

\bibitem{gurevich}  A. Gurevich, Int. J. of Mod. Phys. {\bf 9}, 1045 (1995);
E. H. Brandt, Phys. Rev. Lett. {\bf 76}, 4030 (1996).

\bibitem{burlachkov}  L. Burlachkov, D. Giller, and R. Prozorov, Phys. Rev.
B {\bf 58}, 15067 (1998).

\bibitem{geshkenbein}  V. B. Geshkenbein and A. I. Larkin, Zh. Eksp. Teor.
Fiz. {\bf 95}, 1108 (1989).

\bibitem{zeldov}  E. Zeldov, N. M. Amer, G. Koren, A. Gupta, R. J. Gambino,
and M. W. McElfresh, Phys. Rev. Lett. {\bf 62}, 3093 (1989).

\bibitem{griess}  R. Griessen, A. F. T. Hoekstra, H. H. Wen, G. Doornbos,
and H. G. Schnack, Physica C {\bf 282-287}, 347 (1997); H. H. Wen, A. F. Th.
Hoekstra, R. Griessen, S. L. Yan, L. Fang, and M. S. Si, Phys. Rev. Lett. 
{\bf 79}, 1559 (1997).

\bibitem{abulafia}  Y. Abulafia, A. Shaulov, Y. Wolfus, R. Prozorov, L.
Burlachkov, Y. Yeshurun, D. Majer, E. Zeldov, H. W\"{u}hl, V. B.
Geshkenbein, and V. Vinokur, Phys. Rev. Lett. {\bf 77}, 1596 (1996).

\bibitem{ferrari}  M. J. Ferrari, M. Johnson, F. C. Wellstood, J. Clarke, P.
A. Rosenthal, R. H. Hammond, and M. R. Beasley, Appl. Phys. Lett. {\bf 53},
695 (1988).
\end{references}
\end{document}